\documentclass[12pt]{article}
\usepackage{epsfig}

\usepackage[english]{babel}
\usepackage{amsmath}
\usepackage{amsfonts}
\usepackage{amssymb}
\usepackage{pstricks}
\usepackage{epsf}
\usepackage{psfrag}
\usepackage{epsfig,graphics}

\newcommand{\kb}{\underline{k}}
\newcommand{\rb}{\underline{r}}

\def\si{\tiny}

\def\lsim{\raise0.3ex\hbox{$<$\kern-0.75em\raise-1.1ex\hbox{$\sim$}}}
\def\gsim{\raise0.3ex\hbox{$>$\kern-0.75em\raise-1.1ex\hbox{$\sim$}}}
\def\noi{\noindent}

\def\bei{\begin{itemize}}
\def\ei{\end{itemize}}
\def\beqa{\begin{eqnarray}}
\def\bea{\begin{eqnarray}}
\def\eqa{\end{eqnarray}}
\def\beas{\begin{eqnarray*}}
\def\eeas{\end{eqnarray*}}
\def\beqas{\begin{eqnarray*}}
\def\eqas{\end{eqnarray*}}
\def\beq{\begin{equation}}
\def\eq{\end{equation}}
\def\beqd{\begin{displaymath}}
\def\eeqd{\end{displaymath}}
\def\eqd{\end{displaymath}}

\newcommand{\beeq}[1]{
\marginpar{\small\textsf{#1}}
\begin{equation}\label{#1}}
\newcommand{\eeq}{\end{equation}}
\newcommand{\beea}[1]{
\begin{eqnarray}\label{#1}}
\newcommand{\eea}{\end{eqnarray}}

\def\bef{\begin{frame}}

\def\slashchar#1{\setbox0=\hbox{$#1$}
   \dimen0=\wd0
   \setbox1=\hbox{/} \dimen1=\wd1
   \ifdim\dimen0>\dimen1
      \rlap{\hbox to \dimen0{\hfil/\hfil}}
      #1
   \else
      \rlap{\hbox to \dimen1{\hfil$#1$\hfil}}
      /
   \fi}

\def\kb{\underline{k}}

\newcommand{\fV}{f_{3\,\rho}^V}
\newcommand{\fA}{f_{3\,\rho}^A}

\newcommand{\zV}{\zeta_{3}^V}
\newcommand{\zA}{\zeta_{3}^A}

\def\beq{\begin{equation}}
\def\eeq#1{\label{#1}\end{equation}}
\def\eeqn{\end{equation}}

\def\beqa{\begin{eqnarray}}
\def\eeqa#1{\label{#1}\end{eqnarray}}
\def\eeqan{\end{eqnarray}}

\let\bar=\overbar

\def\Dslash{\not{\hbox{\kern-4pt $D$}}}
\def\dslash{\not{\hbox{\kern-2pt $\del$}}}

\def\msb{{\bar{\ssstyle M \kern -1pt S}}}

\def\Title#1{\begin{center} {\Large {\bf #1} } \end{center}}

\begin{document}

\Title{Hard diffractive processes and non-perturbative matrix elements beyond leading twist: $\rho_T$-meson production}

\bigskip\bigskip

\begin{raggedright}

{\it I.~V.~Anikin$^1$\index{Anikin,I}, D.Yu.~Ivanov$^2$\index{Ivanov,D}, B.~Pire$^3$\index{Pire,B}, L.~Szymanowski$^4$\index{Szymanowski,B}, S.~Wallon$^5$\index{Wallon,S}\\
$1.$ Bogoliubov Laboratory of Theoretical Physics, JINR,
             141980 Dubna, Russia\\
$2.$ Sobolev Institute of Mathematics, 630090 Novossibirsk, Russia\\
$3.$ CPHT, {\'E}cole Polytechnique, CNRS, 91128 Palaiseau Cedex, France\\
$4.$ Soltan Institute for Nuclear Studies, PL-00-681 Warsaw, Poland\\
$5.$ LPT, Universit{\'e} 
Paris-Sud, CNRS, 91405, Orsay, France {\em \ \&}\\
UPMC Univ. Paris 06, facult\'e de physique, 4 place Jussieu, 75252 Paris 05, France}

\end{raggedright}

\section{Introduction}
\label{Sec_Int}
 
Studies of hard exclusive reactions  rely on the factorization properties of the leading twist amplitudes \cite{fact}
for deeply virtual Compton scattering and deep exclusive meson production. The leading twist
distribution amplitude (DA) of a transversally polarized vector meson is
chiral-odd, and hence decouples from hard amplitudes even when another
chiral-odd quantity is involved \cite{DGP} unless in reactions with more than two final
hadrons \cite{IPST}. Thus, transversally polarized $\rho-$meson production is generically governed by
twist 3 contributions for which  a pure collinear
factorization fails  due to the appearance of end-point singularities \cite{MP,AT}.
The meson quark gluon structure within collinear factorization may be described by Distribution Amplitudes (DAs), 
classified in \cite{BB}. On the experimental side, 
in
 photo and electro-production and from moderate to very large energy \cite{expLow,expHigh}, the kinematical analysis of the final $\pi-$meson
  pair allows
to measure the 
$\rho_T-$meson production amplitude,
which is by no means negligible
 and needs to be understood in terms of QCD. Up to now, experimental information comes from electroproduction on a proton
  or nucleus. We will specifically concentrate on the case of  very high energy collisions at colliders.
Future progress in this range may come from real or virtual photon photon collisions \cite{IP,PSW}.

In the literature there are two approaches to the factorization of the
scattering amplitudes in exclusive processes at leading and higher twists. The first approach \cite{APT,AT}, which we will call Light-Cone Collinear Factorization (LCCF), is
the generalization of the Ellis-Furmanski-Petronzio (EFP) method  \cite{EFP} to the exclusive processes, and deals with the factorization in the momentum space around the dominant light-cone
direction.  On the other hand, there exists a Covariant Collinear Factorization (CCF) approach
in coordinate space succesfully applied in \cite{BB} for a systematic
description of DAs of hadrons carrying different
twists.  Although being quite
different and using different DAs, both
approaches can be applied to the description of the same processes.
We have shown  that these two descriptions are equivalent at twist 3 \cite{usSHORT, usLONG}.
We first  
establish a precise vocabulary between objects appearing in the two
approaches. Then
we  calculate
within both methods the impact factor $\gamma^* \to \rho_T$, 
up to twist 3
accuracy, and prove the full consistency between the two results.
The key idea  within LCCF is
the invariance
of the scattering amplitude under rotation of the light-cone  vector $n^\mu$ (conjugated to the light-cone momentum of the partons),
which we call
 $n$-independence condition.  
Combined with the equation of motions (EOMs), this reduces the number of
relevant soft correlators 
to a minimal set. For $\rho$-production up to twist 3,
this reduces a set of 7 DAs to  3 independent DAs which  fully incodes the non-perturbative content of the $\rho$-wave function.


\section{LCCF factorization of exclusive processes} 
 \label{Sec_LCCF}

\subsection{Factorization beyond leading twist}
 \label{SubSec_fact}

The most general form of the  amplitude for the hard exclusive process $A \to \rho \, B$ is,  in
 the momentum representation and in axial
gauge,
%
%
%
%
\begin{eqnarray}
\label{GenAmp}
{\cal A}=
\int d^4\ell \, {\rm tr} \biggl[ H(\ell) \, \Phi (\ell) \biggr]+
\int d^4\ell_1\, d^4\ell_2\, {\rm tr}\biggl[
H_\mu(\ell_1, \ell_2) \, \Phi^{\mu} (\ell_1, \ell_2) \biggr] + \ldots \,,
\end{eqnarray}
where $H$ and $H_\mu$ are  2- and 3-parton coefficient functions,
 respectively. 
In (\ref{GenAmp}), the soft parts are given by the
Fourier-transformed 2- or 3-partons correlators which are matrix elements of non-local operators.
To factorize the amplitude, we  choose the dominant direction around which
we  decompose our relevant momenta and  we Taylor expand the hard part.
Let $p$ and $n$ be a large ``plus" and a small ``minus" light-cone vectors, respectively ($p \cdot n =1$).  Any vector $\ell$ is then expanded as
\begin{eqnarray}
\label{k}
\ell_{i\, \mu} = y_i\,p_\mu  + (\ell_i\cdot p)\, n_\mu + \ell^\perp_{i\,\mu} ,
\quad y_i=\ell_i\cdot n ,
\end{eqnarray}
and  the integration measure in (\ref{GenAmp}) is replaced as
$d^4 \ell_i \longrightarrow d^4 \ell_i \, dy_i \, \delta(y_i-\ell\cdot n) .$
The hard part  $H(\ell)$ is then expanded around
the dominant  ``plus" direction:
\begin{eqnarray}
\label{expand}
H(\ell) = H(y p) + \frac{\partial H(\ell)}{\partial \ell_\alpha} \biggl|_{\ell=y p}\biggr. \,
(\ell-y\,p)_\alpha + \ldots
\end{eqnarray}
where $(\ell-y\,p)_\alpha \approx \ell^\perp_\alpha$ up to twist 3.
To obtain a factorized amplitude, one performs 
an
 integration by parts
to replace  $\ell^\perp_\alpha$ by $\partial^\perp_\alpha$ acting on
the soft correlator.
 This leads to new operators containing
transverse derivatives, such as $\bar \psi \, \partial^\perp \psi $,
 thus requiring
additional DAs
$\Phi^\perp (l)$.
Factorization in the Dirac space is then achieved by 
 Fierz decomposition on a set of relevant $\Gamma$ matrices.
The amplitude is thus factorized as
\bea
\label{GenAmpFac23}
\vspace{-.4cm}{\cal A}&=&
\int\limits_{0}^{1} dy \ {\rm tr} \left[ H_{q \bar{q}}(y) \, \Gamma \right] \, \Phi_{q \bar{q}}^{\Gamma} (y)
+
\int\limits_{0}^{1} dy \ {\rm tr} \left[ H^{\perp\mu}_{q \bar{q}}(y) \, \Gamma \right] \, \Phi^{\perp\Gamma}_{{q \bar{q}}\,\mu} (y) \nonumber
\eea
\bea
&+&\int\limits_{0}^{1} dy_1\, dy_2 \,{\rm tr} \left[ H_{q \bar{q}g}^\mu(y_1,y_2) \, \Gamma \right] \, \Phi^{\Gamma}_{{q \bar{q}g}\,\mu} (y_1,y_2) \,,
\eea
in which the first  (second) line
 corresponds to the 2 (3)-parton contribution  (see Fig.\ref{Fig:Factorized2AND3body}).
\def\si{\footnotesize}
\begin{figure}[h]
\psfrag{rho}[cc][cc]{$\rho$}
\psfrag{k}[cc][cc]{}
\psfrag{rmk}[cc][cc]{}
\psfrag{l}[cc][cc]{$\ell$}
\psfrag{q}[cc][cc]{}
\psfrag{lm}[cc][cc]{}
\psfrag{H}[cc][cc]{\si $H_{q \bar{q}}$}
\psfrag{S}[cc][cc]{\si  $\Phi_{q \bar{q}}$}
\begin{tabular}{ccccc}
\hspace{-.3cm}\epsfig{file=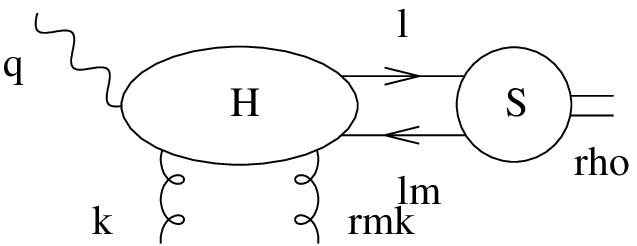,width=4.2cm}
& \hspace{-.5cm} \raisebox{.8cm}{$\longrightarrow $} 
&\hspace{-.6cm}
\psfrag{lm}[cc][cc]{\raisebox{.2cm}{$\quad \,\, \, \Gamma  \ \Gamma$}}
\epsfig{file=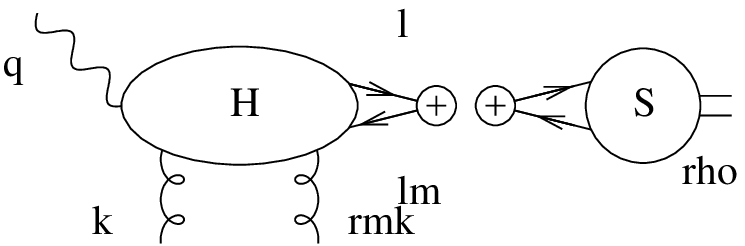,width=5cm}
&\hspace{-.3cm}\raisebox{.8cm}{$+$}
&\hspace{-.1cm}
\psfrag{H}[cc][cc]{{\si $H^\perp_{q \bar{q}}$}}
\psfrag{S}[cc][cc]{\scalebox{1}{\si $\Phi^\perp_{ q \bar{q}}$}}
\psfrag{lm}[cc][cc]{\raisebox{.2cm}{$\quad \,\,\, \, \, \Gamma  \,\, \Gamma$}}
\hspace{-.7cm}
\epsfig{file=FiertzHSqq_rhofact.eps,width=5cm}
\end{tabular}
\psfrag{rho}[cc][cc]{$\rho$}
\psfrag{k}[cc][cc]{}
\psfrag{rmk}[cc][cc]{}
\psfrag{l}[cc][cc]{}
\psfrag{q}[cc][cc]{}
 \psfrag{Hg}[cc][cc]{\si $H_{q \bar{q}g}$}
 \psfrag{Sg}[cc][cc]{\si$\!\! \!\Phi_{q \bar{q}g}$}
 \psfrag{lm}[cc][cc]{}
 \psfrag{H}[cc][cc]{\si $H_{q \bar{q}g}$}
 \psfrag{S}[cc][cc]{\si $\Phi_{q \bar{q}}$}
 \psfrag{S}[cc][cc]{\si $\Phi_{ q \bar{q}g}$}
 \scalebox{1}{\begin{tabular}{ccc}
 \hspace{-0cm}\epsfig{file=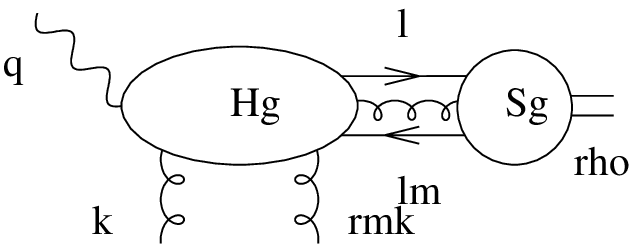,width=4.5cm}
&
\quad \raisebox{.9cm}{$\longrightarrow $} \quad
&
\psfrag{lm}[cc][cc]{\raisebox{.2cm}{$\quad  \, \, \, \, \Gamma  \, \, \Gamma$}}
 \raisebox{.2cm}{\epsfig{file=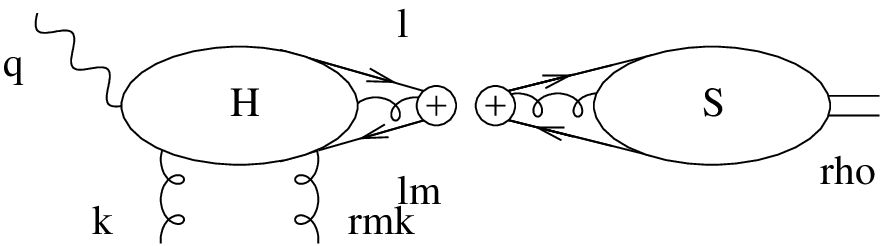,width=5.5cm}}
 \end{tabular}}
\caption{Factorization of 2- (up) and 3-parton (down) contributions in the  example of the $\gamma^* \to \rho$ impact factor.}
\vspace{-.3cm}
\label{Fig:Factorized2AND3body}
\end{figure}
For $\rho$-meson production, the soft parts of ${\cal A}$
 read, with $i\stackrel{\rightarrow}{D}_\mu=i\stackrel{\rightarrow}{\partial}_\mu+\,g \, A_\mu$\,,
\begin{eqnarray}
\label{soft}
\Phi^{\Gamma} (y) &=& \int\limits^{+\infty}_{-\infty} \frac{d\lambda}{2\pi} \, e^{-i\lambda y}
\langle \rho(p) | \bar\psi(\lambda n)\,\Gamma \,\psi(0)| 0 \rangle
\\
\Phi^{\Gamma}_\rho (y_1,y_2)&=& \int\limits^{+\infty}_{-\infty} \frac{d\lambda_1 d\lambda_2}{4\pi^2} \,
e^{-i\lambda y_1\lambda_1-i(y_2-y_1)\lambda_2}
\langle \rho(p) | \bar\psi(\lambda_1 n)\,\Gamma \, i \, \stackrel{\longleftrightarrow}
{D^T_{\rho}}(\lambda_2 n)\, \psi(0)| 0 \rangle \,.
\nonumber
\end{eqnarray}

\subsection{Vacuum--to--rho-meson matrix elements up to twist 3}
\label{SubSec_ParamVacuumRho}



In the LCCF approach,
 the coordinates $z_i$ in the
 parameterizations
have to be fixed by
the light-cone vector $n$. This is in contrast to the CCF approach where $z$ lies on the light cone but does not corresponds
to some fixed light-cone direction.
The transverse polarization of the $\rho-$meson is defined by the conditions
(at twist 3, $p_\rho \sim p$)
\beq
\label{pol_RhoTdef}
e_T \cdot n=e_T \cdot p=0\,.
\eq
Keeping all the terms up to the twist-$3$ order
with the axial (light-like) gauge, $n \cdot A=0$,
the matrix elements of quark-antiquark nonlocal operators
can be written
as (here, $z=\lambda n$ and $\stackrel{{\cal F}_1}{=}$ is
 the Fourier transformation with measure
$\int_{0}^{1}\, dy \,\text{exp}\left[iy\,p\cdot z\right]$)
\begin{eqnarray}
\label{par1v}
&&\langle \rho(p_\rho)|\bar\psi(z)\gamma_{\mu} \psi(0)|0\rangle
\stackrel{{\cal F}_1}{=}
m_\rho\,f_\rho \left[ \varphi_1(y)\, (e^*\cdot n)p_{\mu}+\varphi_3(y)\, e^*_{T\mu}\right],
\\
\label{par1a}
&&\langle \rho(p_\rho)|
\bar\psi(z)\gamma_5\gamma_{\mu} \psi(0) |0\rangle
\stackrel{{\cal F}_1}{=}
m_\rho\,f_\rho \, i\varphi_A(y)\, \varepsilon_{\mu\alpha\beta\delta}\,
e^{*\alpha}_{T}p^{\beta}n^{\delta} \,,
\end{eqnarray}
 where the corresponding flavour matrix has been omitted.
The momentum fraction 

\noindent
$y$ ($\bar y$) corresponds to the quark (antiquark). Denoting 
$\stackrel{\longleftrightarrow}
{\partial_{\rho}}=\frac{1}{2}(\stackrel{\longrightarrow}
{\partial_{\rho}}-\stackrel{\longleftarrow}{\partial_{\rho}})\,,$
the matrix elements of the quark-antiquark operators with transverse derivatives are
\begin{eqnarray}
\label{par1.1v}
&&\langle \rho(p_\rho)|
\bar\psi(z)\gamma_{\mu}
i\stackrel{\longleftrightarrow}
{\partial^T_{\alpha}} \psi(0)|0 \rangle
\stackrel{{\cal F}_1}{=}m_\rho\,f_\rho \,
\varphi_1^T(y) \, p_{\mu} e^*_{T\alpha}
\\
\label{par1.1a}
&&\langle \rho(p_\rho)| \bar\psi(z)\gamma_5\gamma_{\mu}
i\stackrel{\longleftrightarrow}
{\partial^T_{\alpha}} \psi(0) |0\rangle
\stackrel{{\cal F}_1}{=}m_\rho\,f_\rho \,
i\varphi_A^T (y) \, p_{\mu}\, \varepsilon_{\alpha\lambda\beta\delta}\,
e_T^{*\lambda} p^{\beta}\,n^{\delta}\,.
\end{eqnarray}
The matrix elements of quark-gluon nonlocal operators can be
parameterized as\footnote{The symbol $\stackrel{{\cal F}_2}{=}$ means
$
\int\limits_{0}^{1} dy_1 \,\int\limits_{0}^{1} dy_2 \,
\text{exp}\left[ iy_1\,p\cdot z_1+i(y_2-y_1)\,p\cdot z_2 \right] \,.
$}
\begin{eqnarray}
\label{Correlator3BodyV}
&&\langle \rho(p_\rho)|
\bar\psi(z_1)\gamma_{\mu}g A_{\alpha}^T(z_2) \psi(0) |0\rangle
\stackrel{{\cal F}_2}{=}m_\rho\,\fV \,
B(y_1,y_2)\, p_{\mu} e^*_{T\alpha},
\\
\label{Correlator3BodyA}
&&\langle \rho(p_\rho)|
\bar\psi(z_1)\gamma_5\gamma_{\mu} g A_{\alpha}^T(z_2) \psi(0) |0\rangle
\stackrel{{\cal F}_2}{=}m_\rho\,\fA \,
i D(y_1,y_2)\, p_{\mu} \, \varepsilon_{\alpha\lambda\beta\delta} \,
e^{* \, \lambda}_T \, p^{\beta}n^{\delta}\,.
\end{eqnarray}
Note that  $\varphi_1$ corresponds to
the twist-$2$, and
 $B$ and $D$ to the genuine (dynamical) twist-$3$,
while functions $\varphi_3$, $\varphi_A, \varphi_1^T$,
$\varphi_A^T$
contain both parts: kinematical (\`a la
Wandzura-Wilczek, noted WW) twist-$3$ and genuine (dynamical) twist-$3$.


We now recall and rewrite  the original CCF parametrizations of the $\rho$ DAs~\cite{BB},
adapting them to our case when vector meson is produced in the final state, and limiting ourselves to the twist 3 case.
The formula for the axial-vector  correlator is
\beq
\label{BBA}
\langle \rho(p_\rho)|\bar \psi(z) \, [z,\, 0] \, \gamma_\mu \gamma_5 \psi(0)|0\rangle =
\frac{1}{4}f_\rho\,m_\rho\,\varepsilon_\mu^{\;\;\,e^*_T\,p\,z}     \int\limits_0^1\,dy\,e^{iy(p \cdot z)}\,g_\perp^{(a)}(y)\;,
\eq
where we denote
$
\varepsilon_\mu^{\;\;\,e^*_T\,p\,z}= \varepsilon_\mu^{\,\,\,\alpha \beta \gamma} e^*_{T \alpha} \,p_\beta \, z_\gamma\,,$
and in which enters the Wilson line
\beq
\label{defWilson}
[z_1, \, z_2] = P \exp \left[ i g \int\limits^1_0 dt \, (z_1-z_2)_\mu A^\mu(t \,z_1 +(1-t)\,z_2    \right]\,.
\eq
The transverse vector $e_T$ is orthogonal to the light-cone vectors $p$ and $z$, and reads
\beq
\label{pol_Rho}
e_{T \mu}=e_\mu -p_\mu \frac{e \cdot z}{p \cdot z}-z_\mu \frac{e \cdot p}{p \cdot z} \, .
\eq
Thus in the CCF parametrization the 
notion of "transverse" is different with respect to the one of LCCF defined by 
Eq.(\ref{pol_RhoTdef}): as we discuss later in sec.\ref{Sec_Impact}  the coordinate $z$ on the 
light-cone and 
the light-cone vector $n$ point in two different directions.
It is thus useful to rewrite the original CCF 
parametrization in terms of the full meson polarization vector $e$. This is already done for the axial-vector correlator (\ref{BBA})
 since due to the properties of fully antisymmetric tensor 
$\epsilon_{\mu\nu\alpha\beta}$ one can use
$e$ instead of $e_T$  in the r.h.s. of (\ref{BBA}). 
The definition of 2-parton vector  correlator of a
$\rho$-meson  reads
\beq
\label{BBV1}
\langle \rho(p_\rho)|\bar \psi(z) \, [z,\, 0] \, \gamma_\mu  \psi(0)|0\rangle = f_\rho\,m_\rho\int\limits_0^1\,dy\,e^{iy(p\cdot z)}\left[
p_\mu\,\frac{e^*\cdot z}{p\cdot z}\phi_{\parallel}(y) +
e^*_{T\mu}\,g_\perp^{(v)}(y)
  \right]\,,
\eq
which can be rewritten after
 integration by parts in a form which only involve $e$,
\beq
\label{BBV2}
\hspace{0cm}
\langle \rho(p_\rho)|\bar \psi(z) \, [z,\, 0] \, \gamma_\mu  \psi(0)|0\rangle 
= f_\rho\,m_\rho \!\int\limits_0^1 \!\! dy\,e^{iy(p\cdot z)}\!\left[
-i \, p_\mu\,(e^*\cdot z)\,h(y) +
e^*_{\mu}\,g_\perp^{(v)}(y)
  \right],
\eq
with 
$h(y)=\int\limits^y_0 dv \left(\phi_\parallel(v)-g^{(v)}_\perp(v)\right)$
and 
$\bar h(y)=\int\limits^y_0 dv \left(g_3(v)-g^{(v)}_\perp(v)\right) \, .$

\noi
For quark-antiquark-gluon correlators the parametrizations of Ref.\cite{BB} have the forms
\bea
&&
\hspace{-1cm}\langle \rho(p_\rho)|\bar \psi(z)[z,t\, z]\gamma_\alpha g \, G_{\mu\nu}(t\, z)[t\,z,0] \psi(0)|0 \rangle \nonumber \\
&=&
-i p_\alpha [p_\mu e^*_{\perp \nu}-p_\nu e^*_{\perp \mu} ] m_\rho \, \fV \int D \alpha \, V(\alpha_1,\alpha_2)\,
e^{i(p \cdot z)(\alpha_1+t\,\alpha_g)} \, , \label{GV}\\
&&
\hspace{-1cm}\langle \rho(p_\rho)|\bar \psi(z)[z,t\, z]\gamma_\alpha\gamma_5 g \, \tilde G_{\mu\nu}(t\, z)[t\,z,0] \psi(0)|0 \rangle \nonumber \\
&=&
- p_\alpha [p_\mu e^*_{\perp \nu}-p_\nu e^*_{\perp \mu} ] m_\rho \,\fA \int D \alpha \, A(\alpha_1,\alpha_2)\,
e^{i(p \cdot z)(\alpha_1+t\,\alpha_g)} \,,\label{GA}
\eea
where $\alpha_1$, $\alpha_2$, $\alpha_g$ are momentum fractions of quark, antiquark and gluon respectively inside the $\rho-$meson, 
$\int D \alpha =\int\limits^1_0 d\alpha_1\int\limits^1_0 d\alpha_2 \int\limits^1_0 d\alpha_g\,
\delta(1-\alpha_1-\alpha_2-\alpha_g)$
and $\tilde G_{\mu\nu}=-{1\over 2}\epsilon_{\mu\nu\alpha\beta}G^{\alpha\beta}.$
In  the axial gauge $A\cdot n=0$, $n^2=0$,
 the 3-parton correlators thus reads
\begin{eqnarray}
\label{BBVg}
&&\hspace{-1.3cm}\langle \rho(p_\rho)|\bar \psi(z)\gamma_\mu g A_\alpha(tz)\psi(0)|0\rangle \!=\!
-p_\mu \, e^*_{T\alpha}m_\rho \,f^V_{3\rho}\!\int \!
\frac{D \alpha}{\alpha_g} \, e^{i(p\cdot z)(\alpha_1+t\,\alpha_g)}
V(\alpha_1,\alpha_2)\,, \\
\label{BBAg}
&&\hspace{-1.3cm}\langle \rho(p_\rho)|\bar \psi(z)\gamma_\mu\gamma_5 g
A_\alpha(tz)\psi(0)|0\rangle \!=\!
-i p_\mu \frac{\varepsilon_\alpha^{\;\;z\,p\,e^*_T}}{(p\cdot z)}m_\rho \, f^A_{3\rho}\!\int\!
\frac{ D \alpha}{\alpha_g}\, e^{i(p\cdot z)(\alpha_1+t\alpha_g)}
A(\alpha_1,\alpha_2)\,.
\end{eqnarray}


\subsection{Minimal set of DAs and dictionary}

The correlators introduced above are not independent. First, they
are constrained
 by the QCD EOMs for the field
operators entering them (see, for example, \cite{AT}).
In the simplest case of fermionic fields, they follow from the
vanishing of matrix elements
$\langle (i  
{\hat D}(0) \psi(0))_\alpha\, \bar \psi_\beta(z)\rangle = 0$ and
$\langle  \psi_\alpha(0)\, i 
({\hat D}(z)\bar \psi(z))_\beta \rangle
= 0\,$
due to the Dirac equation, then projected on different Fierz structure.
They read, with $\zeta_3^{V(A)} = \frac{f_{3 \, \rho}^{V(A)}}{f_\rho}\,,$
\begin{eqnarray}
\label{em_rho1}
&&\hspace{-1.4cm}\bar{y}_1 \, \varphi_3(y_1) +  \bar{y}_1 \, \varphi_A(y_1)  +  \varphi_1^T(y_1)  +\varphi_A^T(y_1)
=-\!\!\int\limits_{0}^{1} \! \!dy_2 \left[ \zV \, B(y_1,\, y_2) +\zA \, D(y_1,\, y_2) \right], 
\\
&& \hspace{-1.4cm}
\label{em_rho2}
  y_1 \, \varphi_3(y_1) -  y_1 \, \varphi_A(y_1)  -  \varphi_1^T(y_1)  +\varphi_A^T(y_1) =-\!\!\!\int\limits_{0}^{1} \! \!dy_2 \left[ -\zV \, B(y_2,\, y_1) +\zA\, D(y_2,\, y_1) \right] \!\!.
\end{eqnarray}
Second, contrarily to
the light-cone vector $p$ related to the out-going meson momentum,
 the second light-cone vector $n$ (with $p \cdot n=1$), required for the parametrization of the
needed LCCF
correlators,
is arbitrary, and
 the scattering amplitudes should be $n-$independent.
 This  condition expressed at the level of the {\em full amplitude} of any process can be reduced to a set of conditions involving only the soft correlators, and thus the DAs. 
For processes involving $\rho_T$ production up to twist 3 level, we obtained
\begin{eqnarray}
\label{ninV}
&&\hspace{-.5cm}\frac{d}{dy_1}\varphi_1^T(y_1)+\varphi_1(y_1)-\varphi_3(y_1)+\zV\int\limits_0^1\,\frac{dy_2}{y_2-y_1}
 \left( B(y_1,y_2)+B(y_2,y_1) \right)=0\,, 
\\
\label{ninA}
&&\hspace{-.5cm}\frac{d}{dy_1}\varphi_A^T(y_1)-\varphi_A(y_1)+\zA\int\limits_0^1\,\frac{dy_2}{y_2-y_1}
 \left(D(y_1,y_2)+D(y_2,y_1)
   \right)=0 \,.
\end{eqnarray}
The starting point is to exhibit the $n$-dependency of 
the polarization vector for transverse  $\rho$ which enters in the
parametrization of twist 3 correlators, which is
\beq
\label{epsilonRho}
e^{*T}_\mu = e^*_\mu - p_\mu\, e^*\cdot n\,.
\end{equation}
The  $n-$independence condition of the amplitude ${\cal A}$ can thus be written  as
\begin{equation}
\label{nInd}
\frac{d{\cal A}}{dn^\mu}
=0\,, \qquad \mbox{where} \quad  \frac{d}{dn^\mu}= \frac{\partial}{\partial n^\mu} + e^*_\mu\frac{\partial}{\partial (e^* \cdot n)}\,,
\end{equation}
where $e$ denotes now both longitudinal and transverse polarizations.
This will lead to Eqs.(\ref{ninV}, \ref{ninA}) on the
DAs.
Although Eqs.(\ref{ninV}, \ref{ninA}) were derived explicitly in \cite{usLONG} using as a tool the explicit example of the $\gamma^* \to \rho$ impact factor, this proof is independent of the specific process under consideration, and only rely
on general arguments based on Ward identities. For the $\gamma^* \to \rho$ impact factor, 
one needs to consider 2-parton contributions both without  (see Fig.\ref{Fig:NoDer2}) and with  (see Fig.\ref{Fig:Der2}) transverse derivative,
as well as 3-parton contributions (see Figs.\ref{Fig:3Abelian}, \ref{Fig:3NonAbelianOne_Two}). 
The equations (\ref{ninV}, \ref{ninA})
are obtained by
considering the consequence of the $n-$independency on
 the contribution to the $C_F$
color structure\footnote{The $n-$independency condition applied to the $N_c$ structure is automatically satisfied \cite{usLONG}.}. 
To illustrate the idea which is behind this proof, let us consider Eq.(\ref{ninV}),
 which corresponds  to the vector correlator contributions with $C_F$ invariant.
%
In the case of the 3-parton vector correlator (\ref{Correlator3BodyV}), due to (\ref{epsilonRho}) the dependency on
$n$ enters linearly and only through the scalar product $e^* \cdot n\,.$
 Thus, the action on the amplitude of the derivative $d/dn$ defined in (\ref{nInd}) can be extracted by the replacement $e_\alpha^* \to -p_\alpha\,,$ which means in practice that the Feynman rule 
 entering the coupling of the gluon
inside the hard part should be replaced by $-g \, t^a \, \gamma^\alpha \, p_\alpha\,.$
Then, using the Ward identity for the hard part, it reads
\begin{eqnarray}
\label{Ward}
(y_1-y_2){\rm tr} \left[ H^{\rho}_{q \bar{q} g}(y_1,y_2) \, p_\rho
\, \slashchar{p}\right] =
{\rm tr} \left[H_{q \bar{q}}(y_1) \, \slashchar{p}\right]-{\rm tr} \left[H_{q \bar{q}}(y_2) \, \slashchar{p}\right]\,,
\nonumber
\end{eqnarray}
\def\sh{\scriptsize}
which can be seen graphically as

\psfrag{yq}{{\sh $\hspace{-.2cm}y_1$}}
\psfrag{dy}{\raisebox{.04cm}{\sh $\hspace{-.3cm}y_2-y_1$}\raisebox{.6cm}{\sh $\!\!\!\mu$}}
\psfrag{yb}{{\sh $\hspace{-.2cm}1-y_2$}}
\beqa
\psfrag{yq}{{\sh $\hspace{-.2cm}y_1$}}
\psfrag{yb}{{\sh $\hspace{-.2cm}1-y_2$}}
\hspace{-.4cm}
\raisebox{0cm}{$ p_\mu \left[ 
\raisebox{-.8cm}{\includegraphics[width=2.8cm]{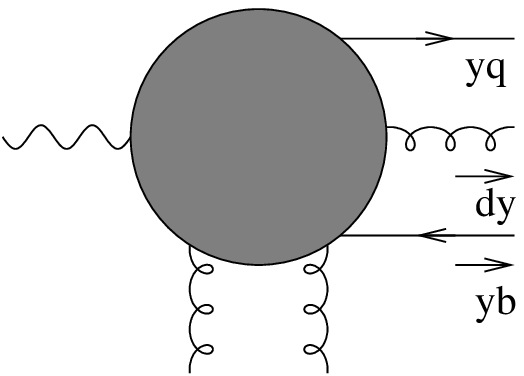}}\hspace{.6cm} \right]$}
\raisebox{0cm}{$=\ \displaystyle \frac{1}{y_1-y_2} \left[
\psfrag{yq}{{\sh $\hspace{-.2cm}y_1$}}
\psfrag{yb}{{\sh $\hspace{-.2cm}1-y_1$}}
\hspace{0.cm}\raisebox{-.8cm}{\includegraphics[width=2.5cm]{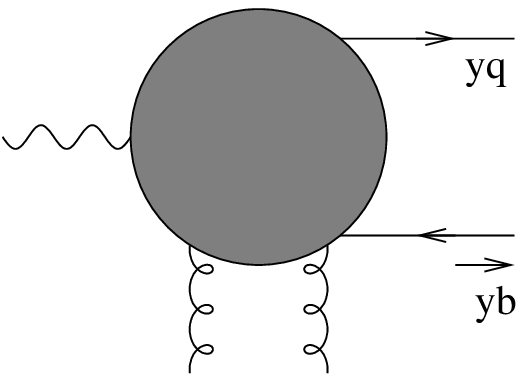}}\hspace{.2cm}
\displaystyle - \hspace{.2cm}
\psfrag{yq}{{\sh $\hspace{-.2cm}y_2$}}
\psfrag{yb}{{\sh $\hspace{-.3cm}1-y_2$}}
\raisebox{-.8cm}{\includegraphics[width=2.5cm]{RWardIF.eps}}\hspace{0.4cm}\right]$}\,.
\label{Fig:WardIF}
\eqa
as is  shown in details in \cite{usLONG}.
Eq.(\ref{Ward}) implies that the 3-parton contribution to the $n-$independency condition can be expressed as the convolution of a 2-parton hard part with the last term of the l.h.s of Eq.(\ref{ninV}).
A similar treatment can be applied to the
2-parton correlators with transverse derivative
whose contributions can be viewed as 3-parton processes with vanishing gluon  momentum. 
This leads to the convolution of the first term of the l.h.s of Eq.(\ref{ninV}) with the {\em same} 2-parton hard part appearing after applying Ward identities to the 3-partons contributions.
The  second term, with $\varphi_1$,  of the l.h.s of Eq.(\ref{ninV})
originates from the 2-parton vector correlator and corresponds to the contribution for the longitudinally
polarized $\rho$ with
$e_L \sim p.$ The third term with $\varphi_3$ corresponds to the contribution of the same correlator for
the polarization vector of  $\rho_T$ written as in Eq.(\ref{epsilonRho}).
To finally get Eq.(\ref{ninV}), we used the fact that each individual term obtained above when expressing the $n-$independency condition involve the
{\em same} 2-parton hard part, convoluted with the Eq.(\ref{ninV}) through an integration over $y_1\,.$ The arguments used above, based on the collinear Ward identity, are clearly independent of the detailled structure of this resulting 2-parton hard part, implying that Eq.(\ref{ninV}) itself should be satisfied.
A similar treatment for axial correlators leads to Eq.(\ref{ninA}), as we have shown in \cite{usLONG}.
 
%
%


Solving
 the 4 equations 
(\ref{em_rho1},\,\ref{em_rho2}, \ref{ninV},\,\ref{ninA}) now
reduces the set of 7 DAs to the set of the 3 independent  DAs $\varphi_1$ (twist 2) and $B,$  $D$ (genuine twist 3).
We write  $\varphi_3(y)$, $\varphi_A(y)$,
$\varphi^T_1(y)$ and  $\varphi^T_A(y)$ generically denoted as $\varphi(y)$
as 
$
\varphi(y)=\varphi^{WW}(y)+\varphi^{gen}(y)$
where $\varphi^{WW}(y)$ and $\varphi^{gen}(y)$ are WW and genuine twist-3 contributions, respectively.
The WW DAs are solutions of
Eqs.~(\ref{em_rho1}, \ref{em_rho2}, \ref{ninV}, \ref{ninA}) with vanishing
$B, \, D$ and read 
\begin{equation}
\label{WWT}
\varphi^{T\;WW}_{A(1)}(y_1)= \frac{1}{2}\left[-\bar y_1
\int\limits_0^{y_1}\,\frac{dv}{\bar v}\varphi_1(v) -(+) \,
 y_1 \int\limits_{y_1}^1\,\frac{dv}{ v}\varphi_1(v)   \right]\;,
\end{equation}
The solution of the set of equations for
  the genuine twist-3 $\varphi^{gen}$ is given in Ref.\cite{usLONG}.



The dictionary betweeen the 3-parton DAs in LCCF and CCF approaches is
\begin{eqnarray}
\label{FullDict}
 B(y_1,\,y_2)&=&-\frac{V(y_1, \, 1-y_2)}{y_2-y_1}\,,  \
D(y_1,\,y_2)=-\frac{A(y_1, \, 1-y_2, \, y_2-y_1)}{y_2-y_1}\,, \nonumber \\
\varphi_1(y)&=&
\phi_{\parallel}(y) \,,
\quad
\varphi_3(y)=
 g_\perp^{(v)}(y) \,, \quad
\varphi_A(y) =
-\frac{1}{4} \, \frac{\partial g_\perp^{(a)}(y)}{\partial y}\,.
\end{eqnarray}

\section{$\gamma^* \to \rho_T$ impact factor up to  twist three accuracy}
\label{Sec_Impact}

\def\li{.14\columnwidth}
\def\si{\hspace{.04cm}}
\def\sci{\hspace{1.7cm}}
\psfrag{u}{\tiny$\hspace{-.3cm}y$}
\psfrag{d}{\tiny$\hspace{-.3cm}y-1$}
\begin{figure}[h]
 \scalebox{1}{\begin{tabular}{cccccc}
 \hspace{-0.1cm}\raisebox{0cm}{\epsfig{file=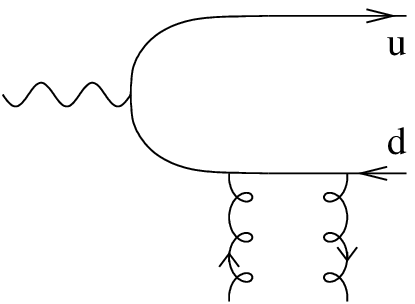,width=\li}} & \si
\psfrag{u}{}
\psfrag{d}{}
 \raisebox{0cm}{\epsfig{file=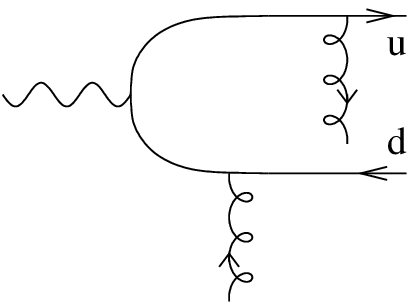,width=\li}} & \si
\psfrag{u}{}
\psfrag{d}{}
\raisebox{0.62cm}{\epsfig{file=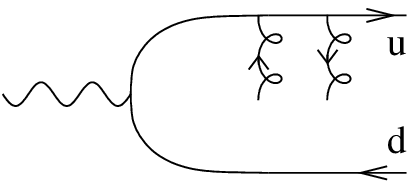,width=\li}}  & \si
\psfrag{u}{}
\psfrag{d}{}
 \raisebox{0cm}{\epsfig{file=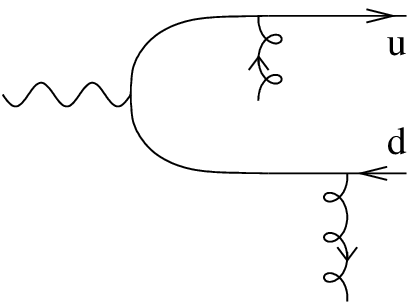,width=\li}}
& \si
\psfrag{u}{}
\psfrag{d}{}
 \raisebox{0cm}{\epsfig{file=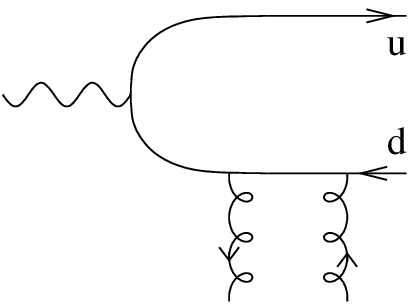,width=\li}}  & \si
\psfrag{u}{}
\psfrag{d}{}
 \raisebox{.62cm}{\epsfig{file=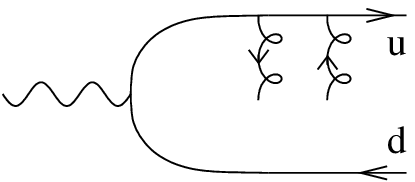,width=\li}} 
 \end{tabular}}
\caption{The 6 hard diagrams attached to the 2-parton correlators, which contribute to the   $\gamma^* \to \rho$ impact factor, with momentum flux of external line along $p_1$ direction.}
\label{Fig:NoDer2}
\end{figure}
\def\li{.14\columnwidth}
\def\si{\hspace{.04cm}}
\def\sci{\hspace{1.7cm}}
\psfrag{i}{}
\psfrag{u}{\tiny$\hspace{-.3cm}y$}
\psfrag{d}{\tiny$\hspace{-.3cm}y-1$}
\psfrag{m}{}
\begin{figure}[h]
 \scalebox{1}{\begin{tabular}{cccccc}
 \hspace{-0.1cm}\raisebox{0cm}{\epsfig{file=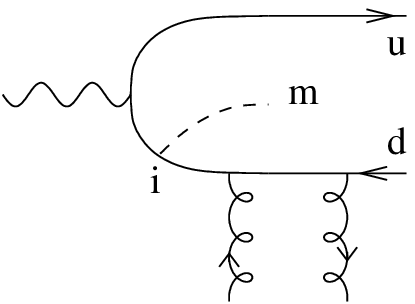,width=\li}} & \si
\psfrag{u}{}
\psfrag{d}{}
 \raisebox{0.06cm}{\epsfig{file=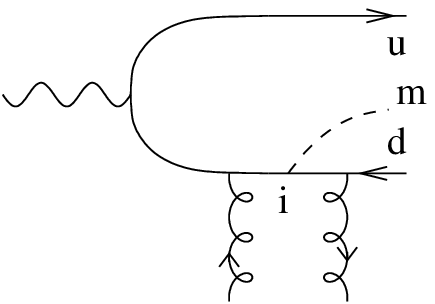,width=\li}} & \si
\psfrag{u}{}
\psfrag{d}{}
\raisebox{0cm}{\epsfig{file=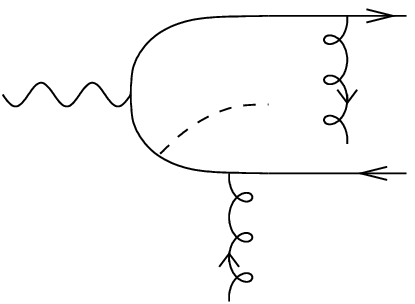,width=\li}}  & \si
\psfrag{u}{}
\psfrag{d}{}
 \raisebox{0cm}{\epsfig{file=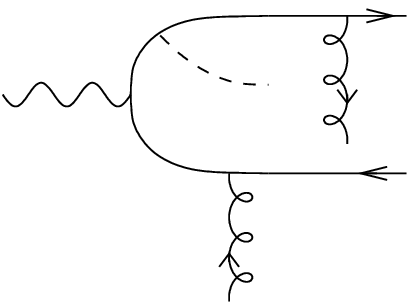,width=\li}}  & \si
\psfrag{u}{}
\psfrag{d}{}
\hspace{0.03cm}\raisebox{.6cm}{\epsfig{file=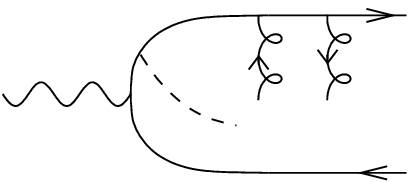,width=\li}} & \si
\psfrag{u}{}
\psfrag{d}{}
 \raisebox{.63cm}{\epsfig{file=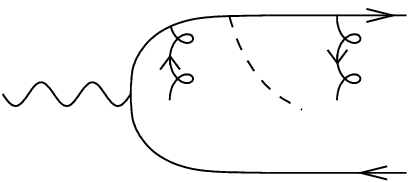,width=\li}} \\
\psfrag{u}{}
\psfrag{d}{}
\raisebox{0cm}{\epsfig{file=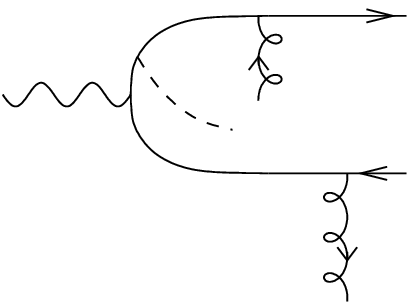,width=\li}}  & \si
\psfrag{u}{}
\psfrag{d}{}
 \raisebox{0cm}{\epsfig{file=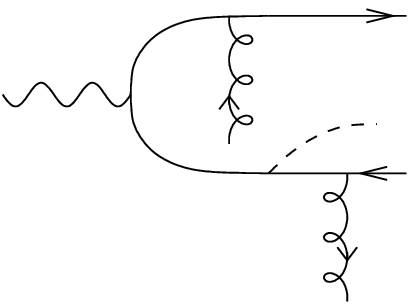,width=\li}}  & \si
\psfrag{u}{}
\psfrag{d}{}
\hspace{0.cm}\raisebox{0cm}{\epsfig{file=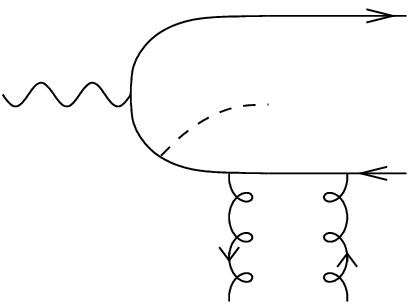,width=\li}} & \si
\psfrag{u}{}
\psfrag{d}{}
 \raisebox{0cm}{\epsfig{file=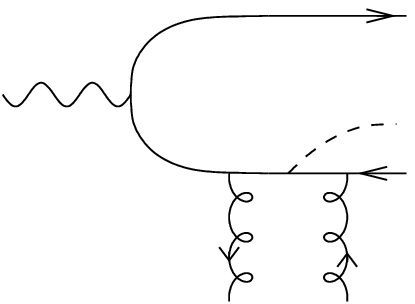,width=\li}} & \si
\psfrag{u}{}
\psfrag{d}{}
\raisebox{.64cm}{\epsfig{file=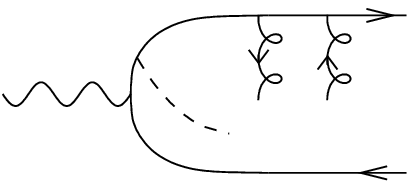,width=\li}}  & \si
\psfrag{u}{}
\psfrag{d}{}
 \raisebox{.64cm}{\epsfig{file=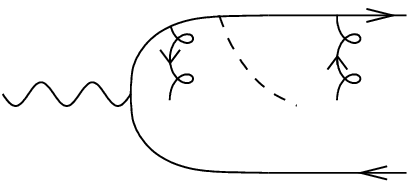,width=\li}}
 \end{tabular}}
\caption{The 12 contributions arising from the first derivative of the 6 hard diagrams attached to the 2-parton correlators, which contribute to the   $\gamma^* \to \rho$ impact factor.}
\label{Fig:Der2}
\end{figure}
\def\si{\hspace{.04cm}}
\begin{figure}[tb]
\psfrag{u}{}
\psfrag{d}{}
\psfrag{m}{}
\psfrag{i}{}
\begin{tabular}{cccccc}
\hspace{-0.3cm} \raisebox{0cm}{\epsfig{file=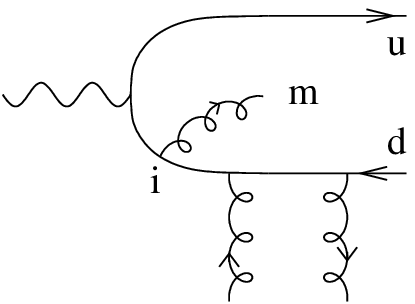,width=\li}}
& \si 
\raisebox{0cm}{\epsfig{file=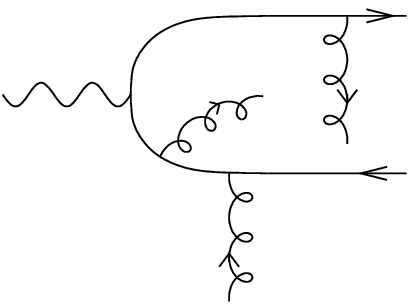,width=\li}}
& \si
\raisebox{0cm}{\epsfig{file=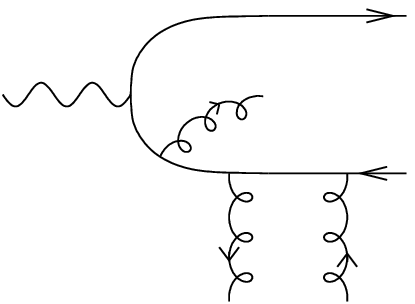,width=\li}}
& \si
\raisebox{.06cm}{\epsfig{file=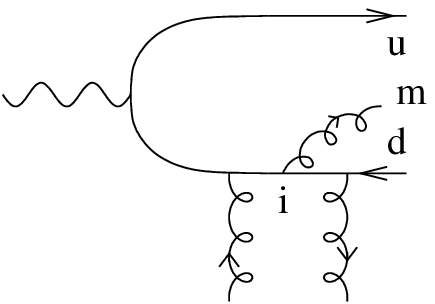,width=\li}}
& \si 
\raisebox{0cm}{\epsfig{file=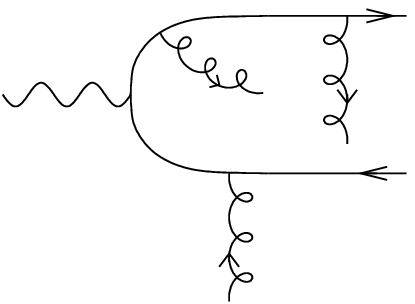,width=\li}} 
& \si
\raisebox{0cm}{\epsfig{file=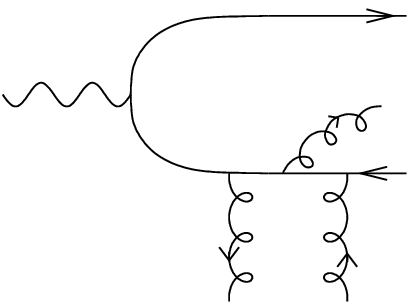,width=\li}} 
\\
\\
\hspace{-0.3cm}\raisebox{0cm}{\epsfig{file=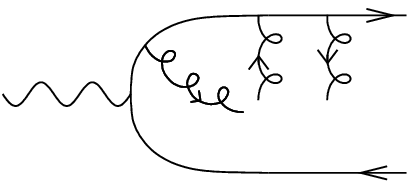,width=\li}}
& \si 
\raisebox{-0.63cm}{\epsfig{file=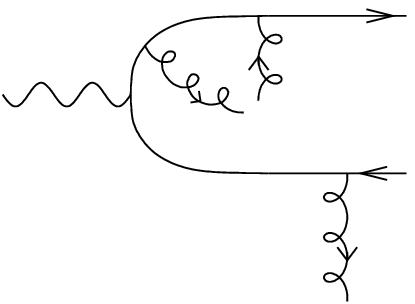,width=\li}}
& \si 
\raisebox{0cm}{\epsfig{file=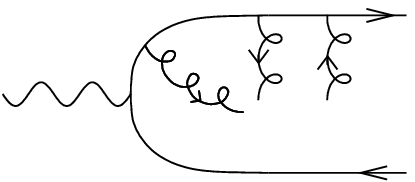,width=\li}}
& \si
\raisebox{0cm}{\epsfig{file=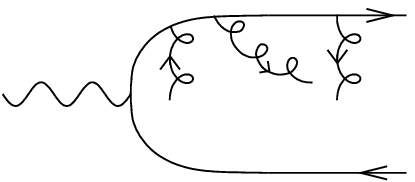,width=\li}}
& \si
\raisebox{-0.63cm}{\epsfig{file=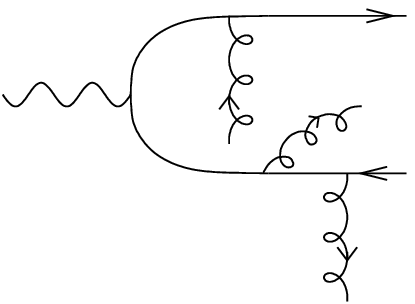,width=\li}} 
& \si
\raisebox{0cm}{\epsfig{file=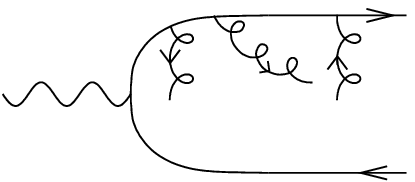,width=\li}}
\end{tabular}
\caption{The 12 ''Abelian`` 
type contributions from  the hard scattering amplitude attached to the 3-parton correlators for the   $\gamma^* \to \rho$ impact factor.}
\label{Fig:3Abelian}
\end{figure}
\def\li{.14\columnwidth}
\def\si{\hspace{.04cm}}
\def\sci{\hspace{1.7cm}}
\psfrag{i}{}
\psfrag{u}{\scalebox{.7}{\tiny$\hspace{-.2cm}y_1$}}
\psfrag{d}{\scalebox{.7}{\tiny$\hspace{-.3cm}-\bar{y}_2$}}
\psfrag{m}{\scalebox{.7}{\tiny$\hspace{-.25cm}y_2-y_1$}}
\begin{figure}[ht]
\begin{tabular}{cccccc}
 \hspace{0.cm}\raisebox{0cm}{\epsfig{file=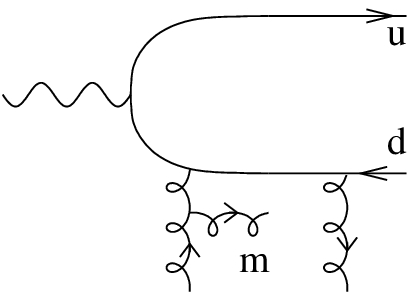,width=\li}} & \si
\psfrag{u}{}
\psfrag{d}{}
\psfrag{m}{}
 \raisebox{0cm}{\epsfig{file=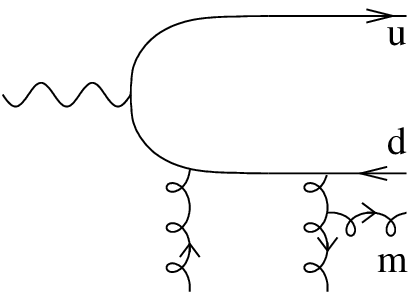,width=\li}} & \si
\psfrag{u}{}
\psfrag{d}{}
\psfrag{m}{}
\raisebox{0cm}{\epsfig{file=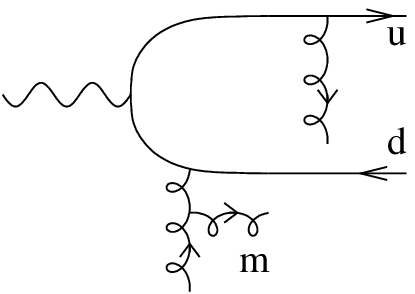,width=\li}}  & \si
\psfrag{u}{}
\psfrag{d}{}
 \psfrag{m}{}
 \raisebox{0.222cm}{\epsfig{file=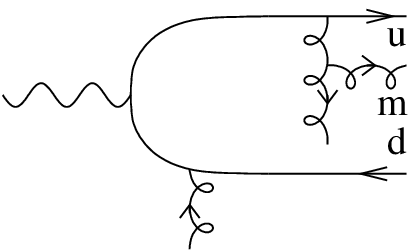,width=\li}}  & \si
\psfrag{u}{}
\psfrag{d}{}
\psfrag{m}{}
\raisebox{0.58cm}{\epsfig{file=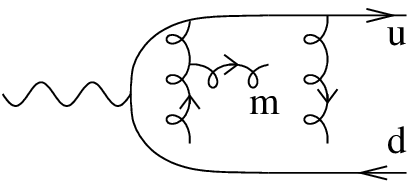,width=\li}} & \si
\psfrag{u}{}
\psfrag{d}{}
\psfrag{m}{}
 \raisebox{0.58cm}{\epsfig{file=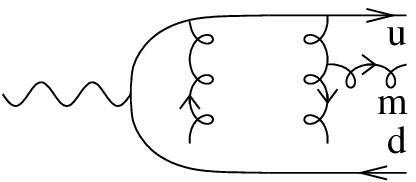,width=\li}} 
\vspace{.1cm}
\\
\psfrag{u}{}
\psfrag{d}{}
\psfrag{m}{}
\raisebox{0.22cm}{\epsfig{file=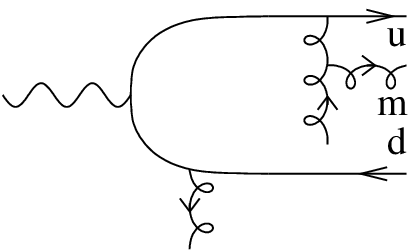,width=\li}}  & \si
\psfrag{u}{}
\psfrag{d}{}
\psfrag{m}{}
 \raisebox{0cm}{\epsfig{file=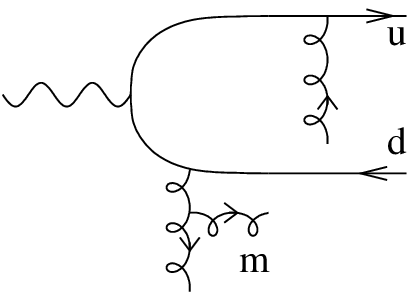,width=\li}}  & \si
\psfrag{u}{}
\psfrag{d}{}
\psfrag{m}{}
\hspace{0.cm}\raisebox{0cm}{\epsfig{file=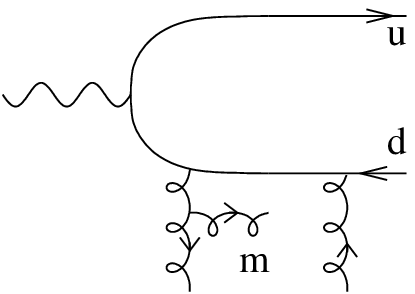,width=\li}} & \si
\psfrag{u}{}
\psfrag{d}{}
\psfrag{m}{}
 \raisebox{0cm}{\epsfig{file=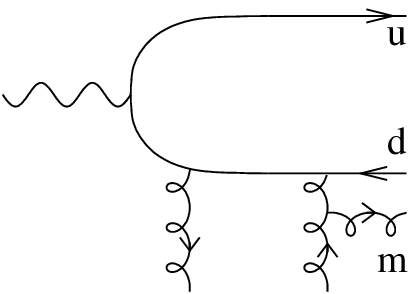,width=\li}} & \si
\psfrag{u}{}
\psfrag{d}{}
\psfrag{m}{}
\raisebox{0.58cm}{\epsfig{file=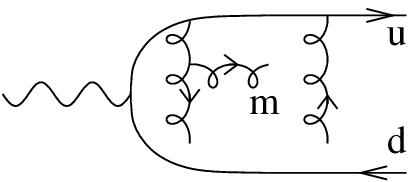,width=\li}}  & \si
\psfrag{u}{}
\psfrag{d}{}
\psfrag{m}{}
 \raisebox{0.58cm}{\epsfig{file=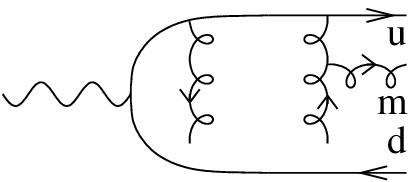,width=\li}}
\vspace{.1cm}
\\
\psfrag{m}{\scalebox{.7}{\tiny$\hspace{-.1cm}y_2-y_1$}}
\raisebox{0cm}{\epsfig{file=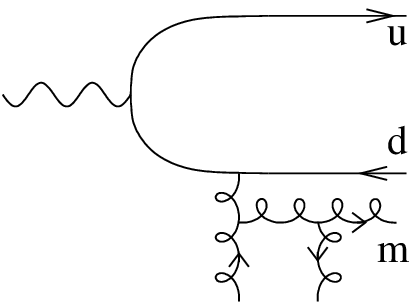,width=\li}} & \si
\psfrag{u}{}
\psfrag{d}{}
\psfrag{m}{}
 \raisebox{0cm}{\epsfig{file=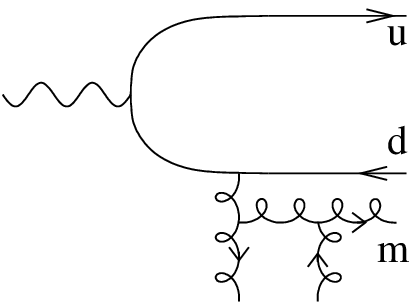,width=\li}} & \si
\psfrag{u}{}
\psfrag{d}{}
\psfrag{m}{}
\raisebox{.62cm}{\epsfig{file=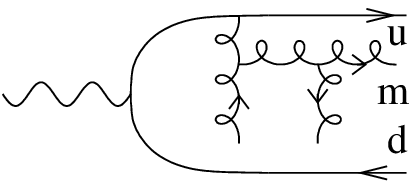,width=\li}}  & \si
\psfrag{u}{}
\psfrag{d}{}
\psfrag{m}{}
 \raisebox{.62cm}{\epsfig{file=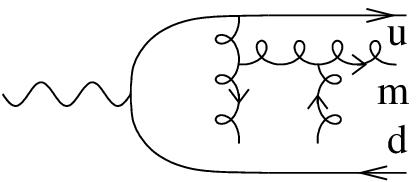,width=\li}} & \si
 \end{tabular}
\caption{The 16 ''non-abelian`` 
(up: one triple gluon vertex, down: two triple gluon vertices) 
contributions 
to the   $\gamma^* \to \rho$ impact factor.}
\label{Fig:3NonAbelianOne_Two}
\end{figure}
%
%
%
%

The $\gamma^* \to \rho $ impact factor enters the description of high energy reactions  in the $k_{T}$-factorization approach, e.g.
 $\gamma^*(q)+N\to \rho_T(p_1)+N$ or
 \begin{eqnarray}
 \label{prgg}
 \gamma^*(q)+\gamma^*(q^\prime)\to \rho_T(p_1)+\rho(p_2)
 \end{eqnarray}
where the virtual photons carry large squared  momenta $q^2=-Q^2$
($q'^2=-Q'^2$) $\gg \Lambda^2_{QCD}$\,, and
the Mandelstam variable $s$ obeys the condition
$s\gg Q^2,\,Q^{\prime\, 2}, -t \simeq \rb^2$. The hard scale which justifies the applicability of
perturbative QCD is set by $Q^2$ and $Q'^2$ and/or by $t.$
Neglecting meson masses, one considers for reaction (\ref{prgg}) the light cone vectors  $p_1$ and $p_2$
as the vector meson momenta  ($2\,p_1\cdot p_2=s$). 
In this Sudakov  basis (transverse euclidian momenta are denoted
 with underlined letters), 
the impact representation of the scattering amplitude for the reaction (\ref{prgg})  is
\begin{eqnarray}
\label{BFKLamforward}
{\cal M}\!=\!\frac{i s}{(2\pi)^2}\!\!
\int\frac{d^2\kb}{\kb^2} \Phi^{ab}_1(\kb,\,\rb-\kb) \!\!
\int\frac{d^2\kb'}{\kb'^2} \Phi^{ab}_2(-\kb',\,-\rb+\kb') \!\!\!\!\!
\int\limits_{\delta-i\infty}^{\delta+i\infty} \frac{d\omega}{2\pi i}
\biggl(\frac{s}{s_0}\biggr)^\omega G_\omega (\kb,\kb',\rb)\,.
\end{eqnarray}
We focus here on the  impact factor $\Phi^{\gamma^* \to \rho}$  of the  subprocess\footnote{The two reggeized
gluons have so-called non-sense polarizations $\varepsilon_1=\varepsilon_2^*=p_2\sqrt{2/s}\,.$}
 $  g(k_1,\varepsilon_{1})+\gamma^*(q)\to g(k_2, \varepsilon_{2})+\rho_T(p_1) \,.$
It is
 the integral of the  S-matrix element
 ${\cal S}^{\gamma^*_T\, g\to\rho_T\, g}_\mu$ with respect to the Sudakov component of the t-channel $k$
momentum along $p_2\,,$ 
 or equivalently the integral of its $\kappa$-channel discontinuity ($\kappa=(q+k_1)^2$) 
\begin{eqnarray}
\label{imfac}
\Phi^{\gamma^*\to\rho}(\kb,\,\rb-\kb)
=
 e^{\gamma^*\mu}\, \frac{1}{2s}\int\limits^{+\infty}_0\frac{d\kappa}{2\pi}
\, \hbox{Disc}_\kappa \,  {\cal S}^{\gamma^*\, g\to\rho\, g}_\mu(\kb,\,\rb-\kb)
\,.
\end{eqnarray}
 Note that within $k_T$-factorization, the description of impact factor for produced hadron described within QCD collinear approach requires a modification of
twist counting due to the off-shellness of the $t-$channel partons. Therefore, when here we say "up to twist 3" we only mean  twist counting from the point of view of the collinear factorization of the produced $\rho-$meson, and not of the whole amplitude, e.g. $\gamma^* \, p \to \rho \, p$ or  $\gamma^* \, \gamma^* \to \rho \, \rho\,.$
We now consider the forward limit for simplicity.
In order to describe the collinear factorization of $\rho$-production inside the impact factor (\ref{imfac}),
we note that the kinematics of the general approach discussed in section \ref{Sec_LCCF} is related to our present kinematics
for the impact factor (\ref{imfac})
by setting $p=p_1$, while a natural choice for $n$ is obtained by setting
$n=p_2/(p_1 \cdot p_2)$ 


We now  compare the
LCCF and CCF approaches, and
 show that they give identical results, when using the dictionary (\ref{FullDict}).
The calculation of the $\gamma^*_L \to \rho_L$  impact factor is standard \cite{ginzburg}. Within LCCF, it receives contribution only from the diagrams with  quark-antiquark
correlators, and it is given by contributions from the $p_\mu$ term  of the correlators (\ref{par1v}) of twist 2.
  It involves the computation of the 6 diagrams of Fig.\ref{Fig:NoDer2}.
We now consider the $\gamma^*_T \to\rho_T$ transition.
The 2-parton
 contribution contains the terms arising from the diagrams of Fig.\ref{Fig:NoDer2}, where
the quark-antiquark correlators have no transverse derivative,
and from the diagrams of Fig.\ref{Fig:Der2}, where the quark-antiquark correlators stand with a
transverse derivatives (denoted with dashed lines).
The contributions of 3-parton correlators are of two types, the first one being of "abelian" type (see Fig.\ref{Fig:3Abelian}) and the second involving non-abelian couplings (see Fig.\ref{Fig:3NonAbelianOne_Two}).
The full result can be decomposed into spin-non-flip and spin-flip
parts, respectively proportional to
$
T_{n.f.}=-(e_\gamma \cdot e^*_T)\,,
$
and 
$T_{f.}=\frac{(e_\gamma \cdot k_\perp)(e^*_T \cdot k_\perp)}{\kb^2}+\frac{(e_\gamma \cdot e^*_T)}{2}\,,
$
and reads
\beq
\label{impactNonFlipFlip}
\Phi^{\gamma^*_T\to\rho_T}(\kb^2)=\Phi_{n.f.}^{\gamma^*_T\to\rho_T}(\kb^2) \, T_{n.f.}+\Phi_{f.}^{\gamma^*_T\to\rho_T}(\kb^2) \, T_{f} \,,
\eq
whose lenghty expressions are given in Ref.\cite{usLONG}.
The gauge invariance of the considered impact factor is checked by the vanishing of our results for $\Phi_{f.}$ and
 $\Phi_{n.f.}$  when $\kb^2=0$.
 The vanishing of the "abelian'', i.e. proportional to $C_F$ part of
$\Phi_{n.f.}$
is particularly subtle since it appears as  a consequence of EOMs
(\ref{em_rho1}, \ref{em_rho2}).
Thus, the expression for the $\gamma^*\to\rho_T$ impact factor has finally
a gauge-invariant form only
provided the genuine twist $3$ contributions have been taken into account.
Finally, we note that 
 end-point
singularities do
not occur here, both in WW approximation and in the full twist-3 order
approximation, due to the $\kb$ regulator specific of $k_T$-factorisation\footnote{This does not preclude the solution of the well known
end-point singularity problem \cite{MP, GK}.}.

We now calculate the impact factor using the CCF parametrization of Ref.\cite{BB} for vector meson
DAs.
 We need to express the impact factor in terms of 
hard coefficient functions and soft parts parametrized by light-cone matrix elements. The standard technique 
here is an operator product expansion on the light cone, $z^2\to 0$, which naturally gives the leading term in the power counting and leads to the described above factorized structure. Unfortunately we do not have an operator definition for an impact factor, and therefore, we have to rely in our actual calculation on the perturbation theory.
However the $z^2\to 0$ limit of any single diagram is given in terms of  
light-cone matrix elements without any Wilson line insertion between the quark and gluon operators, like
$
\langle V(p_V)|\bar \psi(z)\gamma_\mu \psi(0)|0 \rangle$
(we call them as perturbative correlators). 
Actually we need to combine together contributions of quark-antiquark and quark-antiquark gluon diagrams in order to obtain a final gauge invariant result.
At twist 3 level, expanding the Wilson line at first order,  one can show that
\bea
\label{vc-no-wl}
&\langle \rho(p_\rho)|\bar \psi(z)\gamma_\mu \psi(0)|0 \rangle|_{z^2\to0}=& 
\nonumber\\
&f_\rho \, m_\rho \left[-i \,p_\mu (e^* \cdot z)
\int\limits^1_0 dy \, e^{i y (p \cdot z)} (h(y)-\tilde h(y))
+ e^*_{\mu} \int\limits^1_0 dy \, e^{i y (p \cdot z)} g^{(v)}_\perp (z)
\right] \, ,&
\eea    
where  
$\tilde h(y)=\zeta_3^V\int\limits^y_0d\alpha_1\int\limits^{\bar y}_0d\alpha_2\frac{V(\alpha_1,\alpha_2)}{\alpha_g^2}
\, ,
$
with an analogous result for the axial-vector correlator.
Comparing the obtained result (\ref{vc-no-wl}) for the perturbative correlators  with initial parameterizations (\ref{BBV2}) we see that at twist 3-level the net effect of the Wilson line is
just some renormalization of the $h$ function in the case of vector correlator. For the axial-vector we obtain in addition to the function $g^a_\perp$ renormalization a new Lorentz structure (which does not contribute to the impact factor).

Based on the dictionary (\ref{FullDict}) and on the solution of Eqs.(\ref{em_rho1}, \ref{em_rho2}, \ref{ninV}, \ref{ninA}), we got
an exact equivalence between our two LCCF and CCF results, as proven in Ref.\cite{usLONG}.

\section{Conclusion}
\label{Sec_Conclusion}

We compare the momentum space  LCCF  and  the
coordinate space CCF methods,  illustrated here for
 $\rho$-meson production up to twist 3 accuracy. The crucial point 
is the use of Lorentz invariance constraints formulated as the
$n$-independence of the scattering amplitude within LCCF method, which leads to the
necessity of taking into account the contribution of 3-parton correlators.
Our results for the $\gamma^* \rho$ impact factor in both methods are equivalent, based on our dictionary.



\subsection*{Acknowledgments}


 This work is partly supported by the ECO-NET program, contract
18853PJ, the French-Polish scientific agreement Polonium, the grant
ANR-06-JCJC-0084, the RFBR (grants 09-02-01149,
 08-02-00334, 08-02-00896), the grant NSh-1027.2008.2 and
the Polish Grant N202 249235.

\end{document}